\documentclass{iau} 
\usepackage{graphicx}

\title[JD 11.~~LOFT-e: Localisation Of Fast Transients with e-MERLIN] 
{LOFT-e: \\ Localisation Of Fast Transients with e-MERLIN}

\author[C. R. H. Walker \etal]   
{C. R. H. Walker$^1$, R. P. Breton$^1$, P. A. Harrison$^1$, A. Holloway$^1$,\\ M. J. Keith$^1$, M. Kramer$^{1,2}$, M. Malenta$^1$, M. B. Mickaliger$^1$,\\ J. Roy$^3$, T. W. Scragg$^1$
 \and B. W. Stappers$^1$}

\affiliation{$^1$Jodrell Bank Center for Astrophysics, University of Manchester, \\ Alan Turing Building, Oxford Road, Manchester, M13 9PL, UK \\ email: {\tt charles.walker@postgrad.manchester.ac.uk} \\[\affilskip]
$^2$Max-Planck Institute f{\"u}r Radio Astronomy, Auf dem H{\"u}gel 69,
D-53121 Bonn, Germany \\email: {\tt mkramer@mpifr-bonn.mpg.de / michael.kramer@manchester.ac.uk} \\[\affilskip]
$^3$NCRA-TIFR, Pune University Campus, Pune 411007,
India \\email: {\tt jroy@ncra.tifr.res.in}}

\pubyear{2017}
\volume{337}  
\jname{Pulsar Astrophysics -­ The Next 50 Years}
\editors{P. Weltevrede, B.B.P. Perera, L. Levin Preston \& S. Sanidas, eds.}

\begin{document}

\maketitle

\begin{abstract}
The majority of fast radio bursts (FRBs) are poorly localised, hindering their potential scientific yield as galactic, intergalactic, and cosmological probes. LOFT-e, a digital backend for the U.K.'s e-MERLIN seven-telescope interferometer will provide commensal search and real-time detection of FRBs, taking full advantage of its field of view (FoV), sensitivity, and observation time. Upon burst detection, LOFT-e will store raw data offline, enabling the sub-arcsecond localisation provided by e-MERLIN and expanding the pool of localised FRBs. The high-time resolution backend will additionally introduce pulsar observing capabilities to e-MERLIN.
\end{abstract}

\firstsection 
\section{Introduction}

A decade since their initial discovery (\cite[Lorimer \etal\, 2007]{Lori07}),  tens of FRBs (millisecond-duration bursts, presumably of extragalactic origin) exist in literature\footnote{See FRBcat (\cite[Petroff \etal\, 2016]{Petr16}): http://frbcat.org/}. A large future population of FRBs with known redshifts presents exciting opportunities for extragalactic and cosmological astronomy: potentially probing the distribution of matter in the halos of galaxy clusters  (\cite[McQuinn, 2014]{McQu14}), measuring intergalactic magnetic fields, and providing alternative ways to determine cosmological parameters (\cite[Macquart \etal\, 2015]{Macq15}). However, the majority of single-dish-detected FRBs are unconstrained within large (typically arcminute) beams. This places limitations on our ability to determine FRB distances, redshifts, and other parameters, thereby reducing their potential scientific yield. The LOFT-e project is a commensal e-MERLIN backend which will continuously search for FRBs during regular observations, storing suitable candidates for further processing and localisation. This will expand the pool of localised FRBs, taking full advantage of e-MERLIN's field of view (FoV), sensitivity, and observation time.

\section{LOFT-e Overview}

An overview of e-MERLIN's equipment, L-band specifications, and our expectations regarding detection of FRBs with LOFT-e can be found in Table \ref{tab1}. The e-MERLIN array consists of four 25 m dishes (located at Pickmere, Defford, Knockin, Darnhall), one 32 m dish at Cambridge, and two dishes at Jodrell Bank Observatory: the Lovell (76 m) and MkII (32 m equivalent). The facility is capable of observing at L-Band (1.4 GHz), C-band (5 GHz), and K-band (22 GHz). At L-band, the maximum field of view of the array is 30 arcmins (25 m dishes) with a maximum angular resolution of 150 mas and bandwidth of 512 MHz.\\
At maximum capacity, LOFT-e is capable of accepting 12 dual-polarisation data streams from six dishes (2 $\times $64 MHz bandwidth streams per dish). The streams are routed to three computing nodes for real-time processing using two GPUs per node. Upon completion LOFT-e will capture each data stream in parallel in short data buffers, the timeseries data will be digitally channelised, and the resulting filterbanks dedispersed in parallel and searched for single pulses via matched-filter transient searching algorithms. These filterbanks will be subjected to anticoincidence RFI mitigation techniques: pulses found in less than N dishes will be considered RFI and discarded. If significant pulses are found, their raw baseband data will be stored offline for further processing and FRB localisation.\\
When fully deployed, LOFT-e will piggyback standard e-MERLIN observations, allowing for FRB searches without disruption or drain of regular science operations. We expect to search $\sim$1690 hours of L-band data per year, based on 220 days total e-MERLIN on-sky time distributed in a ratio 40:40:20 L:C:K band. From this we expect $\sim$1 FRB detection and localisation per 400 hours, based on the current best-estimate FRB rates (7,000 - 10,000 FRBs sky$^{-1}$ day$^{-1}$ \cite[(Champion \etal\, 2016)]{Cham16}).

\begin{table}
  \begin{center}
  \caption{Overview of e-MERLIN equipment and specifications relevant to LOFT-e, and our expectations regarding FRB detection.}
  \label{tab1}
 {\scriptsize
  \begin{tabular}{|c|c|c|}\hline 
{\bf e-MERLIN Equipment} & {\bf L-Band Specifications} & {\bf LOFT-e Expectations} \\ \hline\hline
  & & \\
{\bf 6 dishes, of}: & {\bf FoV}:& {\bf 1690 hrs in L-band per year}:\\ 
 $6\times$ $25$-$32$ m, & $30$ ' ($25$ m dishes) & from 220 days on sky\\
$1\times$ $76$ m (Lovell) & $\sim 18$ ' (Lovell) & (40:40:20 L:C:K band ratio)\\ 
 & & \\
{\bf 3 receivers}: L-band ($1.4$ GHz), & {\bf Max angular resolution}: 150 mas & {\bf 1 FRB per 400 hrs}: based on\\ 
C-band ($5$ GHz), K-band ($22$ GHz) & {\bf Max bandwidth}: 512 MHz & 7,000-10,000 FRBs sky$^{-1}$day$^{-1}$\\ \hline
  \end{tabular}
  }
 \end{center}
\end{table}

\section{Current Status}

The commissioning of a new e-MERLIN correlator mode to record raw baseband data is complete. Received data is currently being tested to recover bright pulsars such as PSR B0329+54 at L-band and C-band, and to detect giant and single pulses to test searching and anticoincidence techniques. Steps are being taken to enable full automation of the pipeline, and to investigate the RFI environment around e-MERLIN. Future work will include investigation of specific RFI-mitigation techniques such as median average differencing (MAD), completion of the real-time version of the system, and full rollout.

\end{document}